\documentclass[final,smallextended]{svjour2}
\usepackage{empheq}                                   
\usepackage{mathtools,amsfonts,amssymb,amsmath}
\usepackage{times,graphicx,xcolor}
\usepackage{delarray,fancybox}
\usepackage{mathdots}
\usepackage{subfig}
\newcommand{\tc}{\mathtt{t_{c}}}
\newcommand{\ti}{t_{o}}

\newcommand{\tf}{t_{f}}

\newcommand{\cf}{\texttt{1}\hspace{-0.17cm}\texttt{1}}
\newcommand{\nin}{\in\hspace{-0.31cm}/\hspace{0.1cm}}

\title{Heat release by controlled continuous-time Markov jump processes}
\date{}
\author{Paolo Muratore-Ginanneschi \and Carlos Mej\'ia-Monasterio \and  Luca Peliti }
\institute {
Paolo Muratore-Ginanneschi \at University of Helsinki, Department of Mathematics and Statistics,
    P.O.~Box 68 FIN-00014, Helsinki, Finland.
\email{paolo.muratore-ginanneschi@helsinki.fi}
\\
\and Carlos Mej\'ia-Monasterio \at  
Laboratory of Physical Properties, Department of Rural Engineering,
Technical University of Madrid, Av. Complutense s/n, 28040 Madrid, Spain.
\email{carlos.mejia@upm.es}
\\
\and Luca Peliti \at Dipartimento di Scienze Fisiche and Sezione INFN,
Universit\`a “Federico II”, Complesso Monte S. Angelo, I–80126 Napoli, Italy.
\email{peliti@na.infn.it}
}
\begin{document}
\maketitle

\begin{abstract}
  We derive the equations governing the protocols minimizing the heat
  released by a continuous-time Markov jump process on a one-dimensional 
  countable state space during a transition between assigned initial and final 
  probability distributions in a finite time horizon. In particular, we identify
  the hypotheses on the transition rates under which the optimal control
  strategy and the probability distribution of the Markov jump problem
  obey a system of differential equations of
  Hamilton-Jacobi-Bellman-type.  As the state-space mesh tends to zero,
  these equations converge to those satisfied by the diffusion process
  minimizing the heat released in the Langevin formulation of the same
  problem.  We also show that in full analogy with the continuum case,
  heat minimization is equivalent to entropy production minimization.
  Thus, our results may be interpreted as a refined version of the
  second law of thermodynamics.
\end{abstract}

\section{Introduction}
\label{sec:intro}

Molecular motors and more generally nano-machines operate in viscous,
fluctuating environments. It is therefore useful, if not necessary, to
model these systems by means of stochastic processes and to describe
their behavior in terms of probability distributions. Fluctuation
relations
\cite{EvCoMo93,EvSe94,GaCo95,Jar97,Kur98,LeSp99,Cro00,HaSa01} impose
general constraints on these probability distributions which can be
and have been extensively tested experimentally see e.g.
\cite{WaSeMiSeEv02,LiDuSmTiBu02,CaReWaSwSeEv04,TrJaRiCrBuLi04,CoRiJaSmTiBu05,GoPeCiChGa09}.
A unified approach to these results from a theoretical standpoint can
be found in \cite{ChGa07} while a review with emphasis on experimental
aspects and extensive reference to the literature is
\cite{Rit08}. \\
Both in experimental and theoretical investigations of nano-machines
it is crucial to distinguish between fast fluctuating configurational
variables and control parameters, i.e., variables whose state is
determined by external macroscopic sources. For example, in an
experiment following the trajectory of a micron-sized bead immersed in
water and captured in an optical trap, the control parameter could be
the center of the trap measured in the laboratory frame whereas the
configurational
variable is the displacement of the bead \cite{WaSeMiSeEv02,AlRiRi11}. \\
An important observation made in \cite{ScSe07} is that the knowledge
of the control parameters driving a nano-system in a finite-time
transition between two assigned states while minimizing a suitable
non-equilibrium thermodynamical functional (the mean work done on the
system in the examples considered by \cite{ScSe07}), yields
substantial improvements in the measurement from finite-time path
sampling of thermodynamical indicators (e.g. free energy differences)
in both numerical
and experimental studies.   \\
Recent results \cite{AuMeMG11,AuMeMG12,AuGaMeMoMG12} have shown that
the problem posed in \cite{ScSe07} admits a precise mathematical
formulation in the language of optimal control theory see
e.g. \cite{FlemingSoner} and also \cite{vHa07} for a short, brilliant
introduction.  A conspicuous consequence is that for fast processes
described by a Langevin dynamics the minimization of the heat released
or of the work done in a transition between assigned states
can be exactly mapped into Monge-Amp\`ere-Kantorovich optimal transport problems \cite{Villani}.\\
The aim of the present contribution is to extend these results to
transitions described by continuous time Markov jump processes on a
countable state space \cite{Klebaner,KipnisLandim}. Markov jump
processes have been often considered in the study of fluctuation
theorems see e.g. \cite{LeSp99,ImPe07,MaNeWy08,EsBr10,EsBr11} as model
problems for non-equilibrium thermodynamics. The application closest to the scope of 
the present contribution can be found in \cite{EsKaLiBr10}. Namely, the authors of
\cite{EsKaLiBr10} modeled a single-level fermion system interacting with a thermal reservoir 
by a two-state Glauber jump-process with the aim of determining the protocol raising the 
energy level with minimal work done on the system. Rigorous optimal control
theory for jump processes has been developed long ago
\cite{Pli75,DaEl77}. Adapting it to the minimization of the heat
released during a transition between states leads to a formulation in
weak sense of the variational problem similarly to what happens in
stochastic mechanics \cite{GuMo83} and Euclidean quantum mechanics
\cite{Za86}. Heat release minimization is, within our working hypotheses, 
equivalent to entropy production minimization. Thus, our results have a 
straightforward physical interpretation as refined bounds for the second law 
of thermodynamics \cite{AuMeMG12,AuGaMeMoMG12} amenable to direct experimental 
testing \cite{BeArPeCiDiLu12}. In this respect, the aforementioned distinction 
between configurational and control parameters has an immediate, important consequence 
for control theory. Identifying without any further specification the 
jump rates of the process as the controls implies in physical terms \emph{acting at the
fastest possible time scale} of the system. An intuitive and, somewhat trivial, 
consequence is that optimal control will be a jump process. In our view,
a more physically relevant approach is to instead inquire entropy production 
minimization with respect to the broadest set of process parameters 
which lead to smooth, macroscopic optimal control protocols. The mathematical
a-priori condition for this distinction is \emph{coercivity}, a well known concept 
in control theory \cite{FlemingSoner}: the \emph{convexity} of the cost functional with 
respect to the control. Non-coercivity, e.g. linearity of the cost functional with respect
to an optimization parameter brings generically about singular controls. We interpret here 
singular controls as optimization protocols acting on configurational parameters of 
the system and, as such of questionable physical realizability. 
We will argue in the conclusions of this paper that the introduction of 
a control time scale originating from the distinction between 
configurational and control parameters is not a peculiarity of Markov jump process 
but it is also implied in the Langevin dynamics formulation 
\cite{AuMeMG11,AuMeMG12,AuGaMeMoMG12}.
\\
The structure of the paper is as follows. In \S~\ref{sec:jump} we
shortly recall basic properties of Markov jump processes and introduce
the heat functional as a measure of irreversibility in transition
between states.  In order to simplify the notation, we restrict the
attention to a one-dimensional state space.  As noticed in
\cite{AuMeMG12}, the heat optimal control problem is most naturally
formulated in the current velocity formalism \cite{Nelson01} see also
\cite{ChGu10,Gaw11} for application to fluctuation relations.  In
\S~\ref{sec:statement} we state the optimization problem under the
hypothesis that transition rates satisfy a local detailed balance. We
show that in such a case the heat is a convex functional of the
control only if a certain statistical indicator we call ``\emph{reduced
  traffic}'' \cite{MaNeWy08} is bounded. Under this further hypothesis
the optimal control satisfy a system of differential equations of
Hamilton-Jacobi-Bellman type \cite{FlemingSoner}.  In \S~\ref{sec:oc}
we derive these equations which we show in \S~\ref{sec:cl} to recover
in the limit of vanishing lattice spacing the optimal mass transport
equations of \cite{AuMeMG11,AuMeMG12,AuGaMeMoMG12}. Finally, in
\S~\ref{sec:3s} we consider the two and three state dynamics.
For the two state dynamics we show that our optimal protocol corresponds 
to an entropy production lower than the one generated by optimizing the Glauber 
jump process as in \cite{EsKaLiBr10}. This is not surprising 
because using Glauber transition rates is equivalent to fixing the value of 
the reduced traffic. We turn then to the case of a three-state space 
for which we solve numerically the optimal transport equations.

\section{Continuous time Markov jump process}
\label{sec:jump}

Let $\xi_{t}$ a continuous-time Markov jump process taking values on a
one-dimensional countable state space $\mathbb{S}$.  Let
$f(\mathtt{x},t)$ any test function
\begin{eqnarray}
\label{}
f\colon\mathbb{S}\times\mathbb{R}_{+}\mapsto\mathbb{R}
\end{eqnarray}
differentiable at least once with respect to the time variable
$t\in\mathbb{R}_{+}$.  The mean forward derivative of $f$ along the
realizations of $\xi_{t}$ specifies the generator of the process
\begin{eqnarray}
\label{jump:fd}
\lefteqn{
\hspace{-1.0cm}
\lim_{dt\downarrow 0}\mathrm{E}_{\xi_{t}=\mathtt{x}}\left\{\frac{f(\xi_{t+dt},t+dt)-f(\xi_{t},t)}{dt}\right\}
:=(\partial_{t}f+\mathsf{L}f)(\mathtt{x},t)}
\nonumber\\&&
=\partial_{t}f(\mathtt{x},t)+\sum_{\tilde{\mathtt{x}}\in\mathbb{S}}
\left[f(\tilde{\mathtt{x}},t)-f(\mathsf{x},t)\right]\,\mathsf{K}_{t}(\tilde{\mathtt{x}}|\mathtt{x})
\hspace{1.0cm}
\end{eqnarray}
The jump rates $\mathsf{K}_{t}$'s are positive definite (
$\mathsf{K}_{t}(\tilde{\mathtt{x}}|\mathtt{x})\geq 0$ for any pair of
states $\mathtt{x},\tilde{\mathtt{x}}$) and vanishing on the diagonal
( $\mathsf{K}_{t}(\mathtt{x}|\mathtt{x})=0$ for all
$\mathtt{x}\in\mathbb{S}$).  In particular,
$\mathsf{K}_{t}(\tilde{\mathtt{x}}|\mathtt{x})$ denotes the jump rate
from the state $\mathtt{x}$ to $\tilde{\mathtt{x}}$. We allow here the
jump rates to depend upon the time $t$.  The knowledge of the
generator characterizes the process. Namely, the evolution of the
probability distribution
\begin{eqnarray}
\label{}
\mathrm{P}\left(\xi_{t}=\mathtt{x}\right):=m(\mathtt{x},t)
\end{eqnarray}
is governed by the Master equation
\begin{eqnarray}
\label{jump:Master}
\frac{d m}{d t}\left(\mathtt{x},t\right)=\sum_{\tilde{\mathtt{x}}\in\mathbb{S}}
\left\{\mathsf{K}_{t}(\mathtt{x}|\tilde{\mathtt{x}})\,
m\left(\tilde{\mathtt{x}},t\right)-\mathsf{K}_{t}(\tilde{\mathtt{x}}|\mathtt{x})
\,m\left(\mathtt{x},t\right)\right\}
\end{eqnarray}
which for assigned $\mathsf{K}_{t}$'s is a system of non-autonomous
differential equations, infinite dimensional if $\mathbb{S}$ comprises
an infinite number of states.  We require the transition rates to be
sufficiently regular for (\ref{jump:Master}) to admit a unique
solution for any initial probability distribution $m_{o}(\mathtt{x})$
assigned at time $t=\ti$. It is then expedient to introduce the
transition probability $\mathsf{P}_{t,t_{o}}$ as the semi-group
solving
\begin{subequations}
\label{jump:Master2}
\begin{eqnarray}
\label{}
\frac{d \mathsf{P}_{t,\ti}}{d t}\left(\mathtt{x}|\mathtt{x}_{o}\right)=\sum_{\tilde{\mathtt{x}}\in\mathbb{S}}
\left\{\mathsf{K}_{t}(\mathtt{x}|\tilde{\mathtt{x}})\,\mathsf{P}_{t,\ti}\left(\tilde{\mathtt{x}}|\mathtt{x}_{o}\right)
-\mathsf{K}_{t}(\tilde{\mathtt{x}}|\mathtt{x})\,\mathsf{P}_{t,\ti}\left(\mathtt{x}|\mathtt{x}_{o}\right)\right\}
\end{eqnarray}
\begin{eqnarray}
\label{}
\lim_{t\downarrow \ti}\mathsf{P}_{t,\ti}\left(\mathtt{x}|\mathtt{x}_{o}\right)=\cf_{\mathtt{x},\mathtt{x}_{o}}
\end{eqnarray}
\end{subequations}
and to describe the evolution of probability measures as 
\begin{eqnarray}
\label{}
m\left(\mathtt{x},t\right)=\sum_{\tilde{\mathtt{x}}\in\mathbb{S}}
\mathsf{P}_{t,\ti}\left(\mathtt{x}|\tilde{\mathtt{x}}\right)\,m_{o}(\tilde{\mathtt{x}})
:=(\mathsf{P}_{t,\ti}m_{o})(\mathtt{x})
\end{eqnarray}

\subsection{Time-reversal of the probability measure}
\label{sec:jump:tr}

On any finite time horizon $T=\tf-\ti$, under rather general
regularity assumptions on the transition rates $\mathsf{K}_{t}$'s to
any Markov jump process $\xi\equiv\left\{\xi_{t}, t\in[\ti,\tf]\right\}$ 
evolving from an initial probability
distribution $m_{o}(\mathtt{x})$, it is possible to associate a time
reversed process starting at the time-evolved distribution
$m_{f}\left(\mathtt{x}\right)=m\left(\mathtt{x},\tf\right)$. The
corresponding transition probabilities satisfy the time-reversal
condition \cite{Ko36,Nag64}
\begin{eqnarray}
\label{jump:tr:ldb}
\mathsf{P}_{t_{2},t_{1}}\left(\mathtt{x}_{2}|\mathtt{x}_{1}\right)\,m\left(\mathtt{x}_{1},t_{1}\right)
=
\bar{\mathsf{P}}_{t_{1},t_{2}}\left(\mathtt{x}_{1}|\mathtt{x}_{2}\right)\,m\left(\mathtt{x}_{2},t_{2}\right)
\end{eqnarray} 
for any $\ti\leq t_{1}\leq t_{2}\leq \tf$. A straightforward
calculation using (\ref{jump:tr:ldb}) (sketched in
appendix~\ref{ap:mbd}) yields the mean backward derivative of the
process
\begin{eqnarray}
\label{jump:tr:mbd}
\lefteqn{
\lim_{dt\downarrow 0}\mathrm{E}_{\xi_{t}=\mathtt{x}}\left\{\frac{f(\xi_{t},t)-f(\xi_{t-dt},t-dt)}{dt}\right\}
:=(\partial_{t}-\bar{\mathsf{L}}f)(\mathtt{x},t)}
\nonumber\\&&
=(\partial_{t}f)(\mathtt{x},t)
-\sum_{\tilde{\mathtt{x}}\in\mathbb{S}}[f\left(\tilde{\mathtt{x}},t\right)-f\left(\mathtt{x},t\right)]
\frac{\mathsf{K}_{t}(\mathtt{x}|\tilde{\mathtt{x}})\,m(\tilde{\mathtt{x}},t)}{m(\mathtt{x},t)}
\end{eqnarray}
whence we identify the transition rates of the time reversed process 
\begin{eqnarray}
\label{jump:tr:drift}
\bar{\mathsf{K}}_{t}(\tilde{\mathtt{x}}|\mathtt{x}):=
\frac{\mathsf{K}_{t}(\mathtt{x}|\tilde{\mathtt{x}})\,m(\tilde{\mathtt{x}},t)}{m(\mathtt{x},t)}
\end{eqnarray}
It is propaedeutic to our scopes to consider an alternative derivation of 
this classical result. Let us suppose that the process $\xi$ be adapted to the 
\emph{forward} sub-sigma algebra of the natural 
filtration of a Poisson-clock process 
 $\eta\equiv\left\{\eta_{t}, t\in [\ti,\tf]\right\}$ statistically invariant under 
time-reversal and specified by a spatially 
uniform jump-rate $\mathrm{r}_{t}$. Under these hypotheses, Girsanov formula
(see e.g. \cite{KipnisLandim,KoZeSo11} or formula (2.5) of \cite{MaNeWy08}) provides us
with two equivalent representations of the Radon-Nikodym derivative of the 
measure $\mathcal{P}_{\xi}$ of $\xi$ with respect to that $\mathcal{P}_{\eta}$ of $\eta$. 
On the one hand, we can write $d\mathcal{P}_{\xi}/d\mathcal{P}_{\eta}$ in the form of an $\mathcal{P}_{\eta}$-martingale 
with respect to the time $\ti$
\begin{eqnarray}
\label{jump:tr:Girsanov1}
\lefteqn{
\frac{d \mathcal{P}_{\xi }}{d \mathcal{P}_{\eta }}(\eta)
=}
\nonumber\\&&
\exp\left\{-\int_{\ti}^{\tf}dt\,\left[\sum_{\mathtt{x}\in \mathcal{S}}
\mathsf{K}_{t}(\mathtt{x}|\eta_{t})-\mathrm{r}_{t}\right]
+\sum_{ t \in \mathbb{J}(\eta)}\ln \mathsf{K}_{t}(\eta_{t}|\eta_{t^{-}})\right\}
\end{eqnarray}
On the other hand, Girsanov formula yields also the expression for the
Radon-Nykodim derivative of \emph{any} process $\zeta\equiv\left\{\zeta_{t}, t\in[\ti,\tf]\right\}$ and transition rates
$\mathsf{Z}$, absolutely continuous with respect to $\eta$ and, adapted to the its \emph{backward} 
filtration in the form of an $\mathcal{P}_{\eta}$-martingale with respect to the time $\tf$
\begin{eqnarray}
\label{jump:tr:Girsanov2}
\lefteqn{
\frac{d \mathcal{P}_{\zeta }}{d \mathcal{P}_{\eta}}(\eta)
=}
\nonumber\\&&
\exp\left\{-\int_{\ti}^{\tf}dt\left[\sum_{\mathtt{x}\in\mathcal{S}}
\mathsf{Z}_{t}(\mathtt{x}|\eta_{t})-\mathrm{r}_{t}\right]
+\sum_{ t \in \mathbb{J}(\eta)}\ln \mathsf{Z}_{t}(\eta_{t^{-}}|\eta_{t})\right\}
\end{eqnarray}
In writing (\ref{jump:tr:Girsanov1}), (\ref{jump:tr:Girsanov2}) we adhere to 
the convention of considering \emph{c\`adl\`ag}
(equivalently, \emph{corlol}: “continuous on (the) right, limit on (the) left”) paths. 
Furthermore, we denote by $\mathbb{J}(\eta)$ the path-dependent, at most countable set of jumps encountered by 
the realizations of $\eta$. The requirement
\begin{eqnarray}
\label{}
\xi\overset{law}{=}\zeta
\end{eqnarray}
translates then into the condition
\begin{eqnarray}
\label{jump:tr:condition}
\mathrm{m}_{o}(\eta_{\ti})\,\frac{d \mathcal{P}_{\xi }}{d \mathcal{P}_{\eta }}(\eta)
=\mathrm{m}_{f}(\eta_{\tf})\,\frac{d \mathcal{P}_{\zeta }}{d \mathcal{P}_{\eta}}(\eta)
\end{eqnarray}
Upon recalling that increments of 
a smooth function $f$ evaluated along a ``corlol'' step-path, such as the 
realizations of a pure jump process, are amenable to the form
\begin{eqnarray}
\label{}
f(\eta_{t_{2}},t_{2})-f(\eta_{t_{1}},t_{1})=
\int_{t_{1}}^{t_{2}}dt\,\partial_{t}f(\eta_{t},t)+
\sum_{t\in \mathbb{J}(\eta)}[f(\eta_{t},t)-f(\eta_{t^{-}},t)]
\end{eqnarray}
we can solve (\ref{jump:tr:condition}) for the $\mathsf{Z}$'s
to recover (\ref{jump:tr:drift}). Once we determined the analytic expression of 
the time-reversed jump rates, we can use it to construct an auxiliary 
\emph{forward process} $\tilde{\xi}$ with measure  $\mathcal{P}_{\tilde{\xi}}$ absolutely continuous with respect to $\mathcal{P}_{\xi}$. 
We can then apply again Girsanov formula to evaluate the Kullback--Leibler divergence 
\cite{KuLe51} of $\mathcal{P}_{\tilde{\xi}}$ from $\mathcal{P}_{\xi}$ when the two processes start from the same initial data 
at time $\ti$: 
\begin{eqnarray}
\label{jump:tr:KL}
\lefteqn{\mathcal{K}(\mathcal{P}_{\tilde{\xi}}||\mathcal{P}_{\xi}):=
\mathrm{E}^{(\xi)}\ln\frac{d \mathcal{P}_{\xi}}{d \mathcal{P}_{\tilde{\xi}}(\xi)}=
}
\nonumber\\&&
\mathrm{E}^{(\xi)}\left\{-\int_{\ti}^{\tf}dt\,
\sum_{\mathtt{x}\in\mathcal{S}}\left[\mathsf{K}_{t}(\mathtt{x}|\xi_{t})
-\bar{\mathsf{K}}_{t}(\mathtt{x}|\xi_{t})\right]
+ \sum_{t\in\mathbb{J}(\xi)}\ln \frac{\mathsf{K}_{t}(\xi_{t}|\xi_{t_{-}})}
{\bar{\mathsf{K}}_{t}(\xi_{t}|\xi_{t_{-}})}\right\}
\end{eqnarray}  
The notation $\mathrm{E}^{(\xi)}$ emphasizes that the average is with respect to the measure
$\mathcal{P}_{\xi}$. The  Kullback-Leibler divergence (\ref{jump:tr:KL}) provides us with a natural 
probabilistic indicator of the asymmetry between the forward and the backward evolution. 
In particular, we will show in the following section that (\ref{jump:tr:KL}) 
can be identified as the entropy production during a non-equilibrium 
thermodynamic transition from the state $\mathrm{m}_{o}$ to the state
$\mathrm{m}_{f}$ in the time horizon $[\ti,\tf]$. With this goal in view, we observe that 
a straightforward calculation (see appendix~\ref{ap:KL}) allows us
to cast (\ref{jump:tr:KL}) into the form
\begin{eqnarray}
\label{jump:tr:logcost}
\mathcal{K}(\mathcal{P}_{\tilde{\xi}}||\mathcal{P}_{\xi})
=S(\tf)-S(\ti)+\beta\,\mathcal{Q}_{\tf,\ti}
\end{eqnarray}
The first term on the right-hand side coincides with the variation of 
Gibbs-Shannon entropy
\begin{eqnarray}
\label{statement:GB}
S(t)=-\sum_{\mathtt{x}\in\mathbb{S}}(m\,\ln m)(\mathtt{x},t)
\end{eqnarray}
between the states $\mathrm{m}_{o}$ and $\mathrm{m}_{f}$ across the time horizon $[\ti,\tf]$.
The second term is 
\begin{eqnarray}
\label{jump:tr:heat}
\beta\,\mathcal{Q}_{\tf,\ti}=
\int_{\ti}^{\tf}dt\,\sum_{\mathtt{x},\tilde{\mathtt{x}}\in\mathbb{S}}
\ln \frac{\mathsf{K}_{t}(\mathtt{x}|\tilde{\mathtt{x}})}
{\mathsf{K}_{t}(\tilde{\mathtt{x}}|\mathtt{x})}
\mathsf{K}_{t}(\mathtt{x}|\tilde{\mathtt{x}})\,m\left(\tilde{\mathtt{x}},t\right)
\end{eqnarray} 
If we now interpret, following \cite{LeSp99,MaReMo00},
$\mathcal{Q}_{\tf,\ti}$ as the heat released by the process in the
transition between the state $m_{o}$ and $m_{f}$, and $\beta$ with the
inverse of the temperature in units of the Boltzmann constant, the
identification (\ref{jump:tr:heat}) establishes a ``bridge relation''
between the theory of stochastic processes and non-equilibrium
thermodynamics (see \cite{MaNeWy08} and references therein). Two
observations are in order before closing this section.  The first is
that the mathematical hypothesis of absolute continuity guarantees
that the ratios
$\mathsf{K}_{t}(\xi_{t}|\xi_{t_{-}})/\bar{\mathsf{K}}_{t}(\xi_{t}|\xi_{t_{-}})$
in (\ref{jump:tr:KL}) are well defined.  Physically we may identify
this condition with that of ``local detailed balance''
\cite{MaNeWy08}.  The second observation is that the relation between
the Kullback-Leibler divergence (\ref{jump:tr:KL}) and the entropy
production is not limited to Markov jump processes but admits a
straightforward extension to diffusion processes \cite{PMG12}.

\section{Heat release as a ``cost functional''}
\label{sec:statement}

Let us consider now two physical states described by \emph{assigned}
probability distributions $m_{o}$ and $m_{f}$. We are interested in
determining the rates $\mathsf{K}_{t}$'s driving the transition
between $m_{o}$ and $m_{f}$ in a fixed and finite time horizon
$T=\tf-\ti$ such that the heat released in the process has a
minimum. Physical intuition requires that the problem be well-posed if
(\ref{jump:tr:heat}) provides a good definition of the thermodynamical
heat.  This means that the heat must be, as a functional of the
transition rates, bounded from below so to specify a well-defined
``cost function'' for the aforementioned optimal control problem
\cite{FlemingSoner}. This is indeed the case by virtue of a result of
\cite{MaNeWy08}.  Using probability conservation
\begin{eqnarray}
\label{statement:pc}
\sum_{\mathtt{x}\in\mathbb{S}}(\mathsf{L}^{\dagger}\,m)(\mathtt{x},t)=0
\end{eqnarray}
it is possible to couch the integrand in (\ref{jump:tr:heat})
into the form

\begin{eqnarray}
\label{statement:heatrate}
\sum_{\mathtt{x},\tilde{\mathtt{x}}\in\mathbb{S}}
\ln\frac{\mathsf{K}_{t}(\mathtt{x}|\tilde{\mathtt{x}})}{\mathsf{K}_{t}(\tilde{\mathtt{x}}|\mathtt{x})}
\mathsf{K}_{t}(\mathtt{x}|\tilde{\mathtt{x}})
\,m(\tilde{\mathtt{x}},t)=\sigma(t)-\frac{d \,S}{d\,t}(t)
\end{eqnarray}
where
\begin{eqnarray}
\label{statement:epr}
\lefteqn{
\sigma(t):=
}
\nonumber\\&&
\sum_{\mathtt{x},\tilde{\mathtt{x}}\in\mathbb{S}}\frac{\mathsf{K}_{t}(\mathtt{x}|\tilde{\mathtt{x}})\,m(\tilde{\mathtt{x}},t)-
\mathsf{K}_{t}(\tilde{\mathtt{x}}|\mathtt{x})\,m(\mathtt{x},t)}{2}
\ln\frac{\mathsf{K}_{t}(\mathtt{x}|\tilde{\mathtt{x}})\,m(\tilde{\mathtt{x}},t)}
{\mathsf{K}_{t}(\tilde{\mathtt{x}}|\mathtt{x})\,m(\mathtt{x},t)}\,\geq\,0
\end{eqnarray}
can be identified as the entropy production rate. Note that
(\ref{statement:epr}) can be regarded as a consequence of the positive 
definiteness of the Kullback-Leibler divergence (\ref{jump:tr:KL}) with which it 
coincides. The integral version
of (\ref{statement:heatrate})
\begin{eqnarray}
\label{heav:heatwp}
\beta\,\mathcal{Q}=\int_{\ti}^{\tf}dt^{\prime}\,\sigma(t^{\prime})-[S(\tf)-S(\ti)]
\end{eqnarray}
or equivalently the identification
\begin{eqnarray}
\label{statement:seclaw}
S_{\mathrm{Tot.}}(\tf)-S_{\mathrm{Tot.}}(\ti)
:=\beta\,\mathcal{Q}+S(\tf)-S(\ti)\,=\,\int_{\ti}^{\tf}dt^{\prime}\,\sigma(t^{\prime})\,\geq\,0
\end{eqnarray}
has two important consequences:
\begin{itemize}
\item[i] if we interpret $\beta\,\mathcal{Q}_{\tf,\ti}$ as the entropy
  variation at constant temperature $\beta^{-1}$ of the environment during the transformation,
  (\ref{statement:seclaw}) is the expression of the second law of
  thermodynamics \cite{Sei05,ChGa07,MaNeWy08};
\item[ii] for transition between given states, optimal heat control
  \cite{AuMeMG12,AuGaMeMoMG12} reduces effectively to entropy
  production minimization as the boundary conditions fully specify the
  variation of the Gibbs-Shannon entropy.
\end{itemize}

As the logarithm of a strictly positive definite matrix can always be
expressed as the sum of a symmetric and an antisymmetric matrix, we
write the transition rates as
\begin{eqnarray}
\label{statement:rates}
\mathsf{K}_{t}\left(\mathtt{x}|\tilde{\mathtt{x}}\right)=
\mathsf{G}(\mathtt{x},\tilde{\mathtt{x}},t)\,e^{\frac{\mathsf{A}(\mathtt{x},\tilde{\mathtt{x}},t)}{2}}
\end{eqnarray}
where for any fixed time $t$, $\mathsf{G}\colon
\mathbb{S}\times\mathbb{S}\mapsto \mathbb{R}$ is a positive definite,
symmetric function of $\mathtt{x},\tilde{\mathtt{x}}$, such that
$\mathsf{G}\left(\mathtt{x},\mathtt{x},t\right)=0$ for any $x$ and the
function $\mathsf{A}$ is antisymmetric in
$\mathtt{x},\tilde{\mathtt{x}}$.

If we represent the probability distribution of the Markov jump
process into the form
\begin{eqnarray}
\label{}
m(\mathtt{x},t)=e^{R(\mathtt{x},t)}
\end{eqnarray}
then
\begin{eqnarray}
\label{statement:rt}
\mathsf{G}(\mathtt{x},\tilde{\mathtt{x}},t)=
\gamma(\mathtt{x},\tilde{\mathtt{x}},t)\,e^{-\frac{R(\mathtt{x},t)+R(\tilde{\mathtt{x}},t)}{2}}
\end{eqnarray}
where $\gamma$ using the terminology of \cite{MaNeWy08} is the
``\emph{traffic}'' indicator characterizing fluctuations far from
equilibrium of the jump process. At equilibrium
\begin{eqnarray}
\label{}
m_{\star}(\mathtt{x})=e^{R_{\star}(\mathtt{x})} \ ,
\end{eqnarray}
the traffic $\gamma =
(m(\tilde{\mathtt{x}})\mathsf{K}(\mathtt{x}|\tilde{\mathtt{x}}) +
m(\mathtt{x})\mathsf{K} (\tilde{\mathtt{x}}|\mathtt{x}))/2$ measures
the total number of jumps between $\mathtt{x}$ and
$\tilde{\mathtt{x}}$. In view of (\ref{statement:rt}), we
will interpret $\mathsf{G}$ as the state probability distribution
discounted component of the traffic and refer to it as ``\emph{reduced
  traffic}''. 
At equilibrium, the antisymmetric function $\mathsf{A}$ governs
detailed balance relation and reduces to
\begin{eqnarray}
\label{}
\mathsf{A}_{\star}\left(\mathtt{x},\tilde{\mathtt{x}}\right)=R_{\star}(\mathtt{x})-R_{\star}\left(\tilde{\mathtt{x}}\right)
\end{eqnarray}
Out of equilibrium, $\mathsf{A}$ is related to the nonequilibrium driving function
$\mathsf{F}$ as
\begin{eqnarray}
\label{ne:psi}
\mathsf{F}(\mathtt{x},\tilde{\mathtt{x}},t):=\mathsf{A}(\mathtt{x},\tilde{\mathtt{x}},t)-
[R(\mathtt{x},t)-R(\tilde{\mathtt{x}},t)]
\end{eqnarray}
Under these definitions, the relation (\ref{statement:rates}), known
as local detailed balance condition \cite{KaLeSp84,MaNeWy08}, fixes the
values of the symmetric jump rates in terms of the equilibrium density
and the nonequilibrium driving.

In terms of reduced traffic and driving function the total entropy
variation satisfies
\begin{eqnarray}
\label{statement:tev}
\lefteqn{
\mathcal{S}:=S_{\mathrm{Tot.}}(\tf)-S_{\mathrm{Tot.}}(\ti)=
}
\nonumber\\&&
\int_{\ti}^{tf}dt\,
\sum_{\mathtt{x},\tilde{\mathtt{x}}\in\mathbb{S}}\,\mathsf{G}(\mathtt{x},\tilde{\mathtt{x}},t)\,
\mathsf{F}(\mathtt{x},\tilde{\mathtt{x}},t)\,\sinh\frac{\mathsf{F}(\mathtt{x},\tilde{\mathtt{x}},t)}{2}\,
\left[m(\tilde{\mathtt{x}},t)\,m(\mathtt{x},t)\right]^{1/2}
\end{eqnarray}
We will now show that this representation of the total entropy
variation provides a well-defined cost functional to describe the
minimization of the heat release in terms of the control fields
respectively specified by the reduced traffic and the driving
function.  In appendix~\ref{ap:alter} we detail an alternative
formulation to the control problem closer to the approach followed in
\cite{AuMeMG11}.

\section{Optimal control of the total entropy variation}
\label{sec:oc}

Inspection of (\ref{statement:tev}) reveals three important
facts. First, the total entropy variation is a bilinear form in the
probability amplitude
\begin{eqnarray}
\label{oc:pa}
\phi\left(\mathtt{x},t\right)=\sqrt{ m(\mathtt{x},t)}
\end{eqnarray}
evolving by (\ref{jump:Master}) according to the linear law
\begin{subequations}
\begin{eqnarray}
\label{}
\partial_{t}\phi(\mathtt{x},t)=(\mathsf{H}\,\phi)(\mathtt{x},t)
\end{eqnarray}
\begin{eqnarray}
\label{}
(\mathsf{H}\,\phi)(\mathtt{x},t):=\sum_{\tilde{\mathtt{x}}\in\mathbb{S}}\mathsf{G}(\mathtt{x},\tilde{\mathtt{x}},t)
\sinh\frac{\mathsf{F}(\mathtt{x},\tilde{\mathtt{x}},t)}{2}\phi(\tilde{\mathtt{x}},t)\,
\end{eqnarray}
\end{subequations}
The ``Hamiltonian'' operator $\mathsf{H}\colon\mathbb{S}\times\mathbb{S}\times\mathbb{R}_{+}\mapsto\mathbb{R}$ 
is for any fixed time $t\in\mathbb{R}_{+}$ anti-symmetric under states $(\mathtt{x},\tilde{\mathtt{x}})$
permutation  
\begin{eqnarray}
\label{}
\mathsf{H}^{\dagger}=-\mathsf{H}
\end{eqnarray}
thus enforcing at any time probability conservation:
\begin{eqnarray}
\label{oc:pc}
\sum_{\mathtt{x}\in\mathbb{S}}\phi^{2}(\mathtt{x},t)=1\hspace{1.0cm}\forall\,t
\end{eqnarray}
As a consequence, the optimal control problem can be treated in full
analogy with the variational techniques described in \cite{Za86} in
the context of Euclidean quantum mechanics. The second fact is that
(\ref{statement:tev}) is linear in the reduced traffic.  Cost
functionals linear in the control are known in general to lead to
singular solutions i.e. not satisfying smooth (partial) differential
equations \cite{FlemingSoner}.  The third fact is that the total
entropy variation is a convex functional of the current potential
$\mathsf{F}$. We expect therefore that if the reduced traffic is
bounded from below or simply constrained to a fixed value, the optimal
control problem admits a unique solution with $\mathsf{F}$ specified
by a differential equation of Hamilton-Jacobi-Bellman type
\cite{FlemingSoner,vHa07}. Using (\ref{oc:pa}), we write the cost
functional specified by the heat release between two given states as
\begin{subequations}
\label{oc:cost}
\begin{eqnarray}
\label{oc:costfun}
\mathcal{S}:=
\int_{t_{o}}^{\tf}dt\,\sum_{\mathtt{x},\tilde{\mathtt{x}}\in\mathbb{S}} 
\phi(\mathtt{x},t)\,\mathsf{U}(\mathtt{x},\tilde{\mathtt{x}},t)
\,\phi(\tilde{\mathtt{x}},t)\equiv \int_{t_{o}}^{\tf}dt\,\boldsymbol{\phi}\cdot \mathsf{U}\cdot\boldsymbol{\phi}
\end{eqnarray}
\begin{eqnarray}
\label{oc:costlag}
\mathsf{U}(\mathtt{x},\tilde{\mathtt{x}},t):=\mathsf{G}(\mathtt{x},\tilde{\mathtt{x}},t)
\,\mathsf{F}(\mathtt{x},\tilde{\mathtt{x}},t)\,\sinh\frac{\mathsf{F}(\mathtt{x},\tilde{\mathtt{x}},t)}{2}\,
\end{eqnarray}
\end{subequations}
In (\ref{oc:cost}), we regard the reduced traffic $\mathsf{G}$ and the
driving function $\mathsf{F}$ as independent controls only restricted
by the requirement that the time boundary conditions on the
probability amplitudes be satisfied. As a consequence the total
variation of the cost functional decomposes into
\begin{eqnarray}
\label{}
\mathcal{S}^{\prime}=\mathcal{S}^{\prime}_{\mathsf{G}}+\mathcal{S}^{\prime}_{\mathsf{F}}
\end{eqnarray}
where
\begin{eqnarray}
\label{oc:ve1}
\mathcal{S}^{\prime}_{\mathsf{X}}=\int_{\ti}^{\tf}dt\,\left\{\boldsymbol{\phi}\cdot\mathsf{U}^{\prime}_{\mathsf{X}}\cdot
\boldsymbol{\phi}+\boldsymbol{\phi}^{\prime}_{\mathsf{X}}\cdot\mathsf{U}\cdot
\boldsymbol{\phi}+\boldsymbol{\phi}\cdot\mathsf{U}\cdot
\boldsymbol{\phi}^{\prime}_{\mathsf{X}}\right\}
\end{eqnarray} 
for $\mathsf{X}=(\mathsf{G},\mathsf{F})$. In (\ref{oc:ve1}) and in
what follows we define 
the first variation as
\begin{eqnarray}
\label{}
\mathsf{O}^{\prime}_{\mathsf{X}}:=\sum_{\mathtt{y},\tilde{\mathtt{y}}\in\mathbb{S}}
\mathsf{X}^{\prime}\left(\mathtt{y},\tilde{\mathtt{y}},t\right)\frac{\partial\,\mathsf{O}}
{\partial\mathsf{X}\left(\mathtt{y},\tilde{\mathtt{y}},t\right)}
\end{eqnarray}
we also assume that $\mathsf{X}^{\prime}$ has the same parity of
$\mathsf{X}$ under permutation of its state space arguments.
Bellman principle states that the optimal Markov control corresponds
to the stationary variation of a local functional $\mathsf{J}$, named
the \emph{value function}, of the stochastic process.  We will now
show that for heat release minimization the interpretation of Bellman
principle in a \emph{weak sense} similar to \cite{GuMo83} yields the
optimal control strategy of physical interest. We start by defining
the value function as the solution of
\begin{eqnarray}
\label{oc:value}
\partial_{t}\mathsf{J}\left(\mathtt{x},\tilde{\mathtt{x}},t\right)=
[\mathsf{H}\,,\mathsf{J}]\left(\mathtt{x},\tilde{\mathtt{x}},t\right)-
\mathsf{U}\left(\mathtt{x},\tilde{\mathtt{x}},t\right)
\end{eqnarray}
where as usual
\begin{eqnarray}
\label{}
[\mathsf{H}\,,\mathsf{J}]\left(\mathtt{x},\tilde{\mathtt{x}},t\right)=\sum_{\mathtt{y}\in\mathbb{S}}\left\{
\mathsf{H}\left(\mathtt{x},\mathtt{y},t\right)\mathsf{J}\left(\mathtt{y},\tilde{\mathtt{x}},t\right)
-\mathsf{J}\left(\mathtt{x},\mathtt{y},t\right)\mathsf{H}\left(\mathtt{y},\tilde{\mathtt{x}},t\right)
\right\}
\end{eqnarray}
We require $\mathsf{J}$ to satisfy the final boundary conditions
\begin{eqnarray}
\label{oc:valuebc}
\mathsf{J}_{\mathsf{F}}^{\prime}\left(\mathtt{x},\tilde{\mathtt{x}},\tf\right)
=\mathsf{J}_{\mathsf{G}}^{\prime}\left(\mathtt{x},\tilde{\mathtt{x}},\ti\right)=0
\end{eqnarray}
These conditions stem from the fact that admissible controls are only
those driving the Markov process between assigned probability
distributions at the ends of the control horizon. For this reason, we
require $\mathsf{J}(\cdot,\tf)$ to be a pure functional of
$\boldsymbol{\phi}_{f}$ whence (\ref{oc:valuebc}) follows.  An
immediate consequence of (\ref{oc:value}) is
\begin{eqnarray}
\label{}
\mathcal{S}=(\boldsymbol{\phi}\cdot\mathsf{J}\cdot\boldsymbol{\phi})(\ti)-(\boldsymbol{\phi}\cdot\mathsf{J}\cdot\boldsymbol{\phi})(\tf)
\end{eqnarray}
Using (\ref{oc:value}) and performing a time-integration by parts we can recast (\ref{oc:ve1}) 
into the form
\begin{eqnarray}
\label{oc:ve2}
\lefteqn{
\mathcal{F}^{\prime}_{\mathsf{X}}=(\boldsymbol{\phi}\cdot\mathsf{J}_{\mathsf{X}}^{\prime}\cdot\boldsymbol{\phi})(\ti)
}
\nonumber\\&&
+\int_{\ti}^{\tf}dt\,\boldsymbol{\phi}\cdot(\partial_{t}\mathsf{J}-[\mathsf{H}\,,\mathsf{J}]+\mathsf{U})_{\mathsf{X}}^{\prime}\cdot\boldsymbol{\phi}
=(\boldsymbol{\phi}\cdot\mathsf{J}_{\mathsf{X}}^{\prime}\cdot\boldsymbol{\phi})(\ti)
\end{eqnarray}
we can now discriminate between three different cases.

\subsection{Bellman principle in strong sense}

Interpreted in strong sense, Bellman principle implies
\begin{eqnarray}
\label{}
\frac{\partial \mathsf{J}(\mathtt{x},t)}{\partial \mathsf{X}(\mathtt{y},\tilde{\mathtt{y}},t)}
=0\hspace{1.0cm}\forall \,t\,\in\,[\ti,\tf]
\end{eqnarray} 
The stationarity conditions take therefore the general form
\begin{eqnarray}
\label{}
0=\left[\mathsf{H}_{\mathsf{X}}^{\prime}\,,\mathsf{J}\right]-\mathsf{U}_{\mathsf{X}}^{\prime}
\end{eqnarray}
which reduce to the system of equations
\begin{subequations}
\label{ocss:sc}
\begin{eqnarray}
\label{ocss:sc1}
0=\left\{\delta_{\mathtt{y},\mathtt{x}}\mathsf{J}(\tilde{\mathtt{y}},\tilde{\mathtt{x}},t)
-\delta_{\tilde{\mathtt{y}},\tilde{\mathtt{x}}}\mathsf{J}(\mathtt{x},\mathtt{y},t)
-\delta_{\mathtt{y},\mathtt{x}}\delta_{\tilde{\mathtt{y}},\tilde{\mathtt{x}}}\mathsf{F}(\mathtt{x},\tilde{\mathtt{x}},t)\right\}
\sinh\frac{\mathsf{F}(\mathtt{y},\tilde{\mathtt{y}},t)}{2}
\end{eqnarray}
\begin{eqnarray}
\label{ocss:sc2}
\lefteqn{
0=\frac{\delta_{\mathtt{y},\mathtt{x}}\mathsf{J}(\tilde{\mathtt{y}},\tilde{\mathtt{x}},t)
-\delta_{\tilde{\mathtt{y}},\tilde{\mathtt{x}}}\mathsf{J}(\mathtt{x},\mathtt{y},t)}{2}
\mathsf{G}(\mathtt{y},\tilde{\mathtt{y}},t)\cosh\frac{\mathsf{F}(\mathtt{y},\tilde{\mathtt{y}},t)}{2}
}
\nonumber\\&&
-\delta_{\mathtt{y},\mathtt{x}}\delta_{\tilde{\mathtt{y}},\tilde{\mathtt{x}}}
\left\{
\sinh\frac{\mathsf{F}(\mathtt{x},\tilde{\mathtt{x}},t)}{2}+\frac{\mathsf{F}(\mathtt{x},\tilde{\mathtt{x}},t)}{2}
\cosh\frac{\mathsf{F}(\mathtt{x},\tilde{\mathtt{x}},t)}{2}\right\}\mathsf{G}(\mathtt{y},\tilde{\mathtt{y}},t)
\end{eqnarray}
\end{subequations}
Regarding $\mathsf{J}$ as a square matrix in the state-space
variables, (\ref{ocss:sc1}) imposes that only diagonal element be
non-vanishing. The condition
\begin{eqnarray}
\label{}
\mathsf{J}(\tilde{\mathtt{x}},\tilde{\mathtt{x}},t)
-\mathsf{J}(\mathtt{x},\mathtt{x},t)-\mathsf{F}(\mathtt{x},\tilde{\mathtt{x}},t)=0
\end{eqnarray}
is, however, consistent with (\ref{ocss:sc2}) for $\mathsf{G}>0$ only if 
\begin{eqnarray}
\label{oc:scsol}
\mathsf{F}(\mathtt{x},\tilde{\mathtt{x}},t)=0\hspace{1.0cm} \forall\,\mathtt{x},\tilde{\mathtt{x}}\in\mathbb{S}
\end{eqnarray}
The conclusion is that an optimal control strategy compatible with the
boundary conditions must be singular. 
This means that the entropy production attains its infimum on a protocol
instantaneously switching at any time during the control horizon 
from a state of equilibrium with the initial condition $\mathrm{m}_{o}$ 
to a state in equilibrium with the final $\mathrm{m}_{f}$. 
We won't delve any deeper here on singular control but we
will return to it in the conclusions.

\subsection{Bellman principle in weak sense}

The equality
\begin{eqnarray}
\label{}
(\boldsymbol{\phi}\cdot\mathsf{J}_{\mathsf{X}}^{\prime}\cdot\boldsymbol{\phi})(\ti)
=\int_{\ti}^{\tf}dt\,(\boldsymbol{\phi}\cdot\left(\mathsf{U}_{\mathsf{X}}^{\prime}-[\mathsf{H}_{}^{\prime}\,,\mathsf{J}]\,\right)
\cdot\boldsymbol{\phi})(t)
\end{eqnarray}
yields the weak-sense stationarity conditions
\begin{subequations}
\label{oc:scws}
\begin{eqnarray}
\label{oc:scws1}
\lefteqn{
0=\phi(\mathtt{y},t)\mathsf{F}(\mathtt{y},\tilde{\mathtt{y}},t)\phi(\tilde{\mathtt{y}},t)
}
\nonumber\\&&
-\sum_{\mathtt{x}\in\mathbb{S}}\left\{\phi(\mathtt{y},t)\mathsf{J}(\tilde{\mathtt{y}},\mathtt{x},t)\phi(\mathtt{x},t)
-\phi(\mathtt{x},t)\mathsf{J}(\mathtt{x},\mathtt{y},t)\phi(\tilde{\mathtt{y}},t)\right\}
\end{eqnarray}
\begin{eqnarray}
\label{oc:scws2}
\lefteqn{
0=\sum_{\mathtt{x}\in\mathbb{S}}\left[\phi(\mathtt{y},t)\mathsf{J}(\tilde{\mathtt{y}},\mathtt{x},t)\phi(\mathtt{x},t)
-\phi(\mathtt{x},t)\mathsf{J}(\mathtt{x},\mathtt{y},t)\,\phi(\tilde{\mathtt{y}},t)\right]
}
\nonumber\\&&
-\phi(\mathtt{y},t)\,\phi(\tilde{\mathtt{y}},t)
\left\{2\,
\tanh\frac{\mathsf{F}(\mathtt{y},\tilde{\mathtt{y}},t)}{2}+\mathsf{F}(\mathtt{y},\tilde{\mathtt{y}},t)
\right\}\hspace{1.2cm}
\end{eqnarray}
\end{subequations}
The conditions are satisfied if $\mathsf{J}$ is symmetric under
permutation of the state variables. Furthermore, if we define the
auxiliary field $B$
\begin{eqnarray}
\label{}
\phi(\mathtt{y},t)\,B(\mathtt{y},t)
=\sum_{\mathtt{x}\in\mathbb{S}}\mathsf{J}(\mathtt{y},\mathtt{x},t)\phi(\mathtt{x},t)
\end{eqnarray} 
the pair (\ref{oc:scws}) simplifies to
\begin{subequations}
\label{oc:scwsv2}
\begin{eqnarray}
\label{oc:scwsv21}
0=B(\mathtt{y},t)-B(\tilde{\mathtt{y}},t)+\mathsf{F}(\mathtt{y},\tilde{\mathtt{y}},t)
\end{eqnarray}
\begin{eqnarray}
\label{oc:scwsv22}
0=B(\mathtt{y},t)-B(\tilde{\mathtt{y}},t)+
2\,\tanh\frac{\mathsf{F}(\mathtt{y},\tilde{\mathtt{y}},t)}{2}
+\mathsf{F}(\mathtt{y},\tilde{\mathtt{y}},t)
\end{eqnarray}
\end{subequations}

\subsection{Unbounded reduced traffic leads to singular control}

As in the strong sense case, if the reduced traffic can be varied without 
any lower bound (\ref{oc:scwsv2}) is satisfied by (\ref{oc:scsol}).
The optimal control strategy is again singular.

\subsection{A constraint on reduced traffic leads to an Hamilton-Jacobi-Bellman equations}

A more interesting and presumably more physically relevant situation
occurs when the reduced traffic is bounded from below by a given
constraint. In such a case we expect the control to consists of a
singular part pushing $\mathsf{G}$ to its lower bound and then of a
smoother part, which minimizes the entropy variation versus
$\mathsf{F}$ for fixed $\mathsf{G}$. The limit case is when
$\mathsf{G}$ is constrained from the very beginning to a given value
e.g.
\begin{eqnarray}
\label{}
\mathsf{G}(\mathtt{x},\tilde{\mathtt{x}},t)=\frac{1-\delta_{\mathtt{x},\tilde{\mathtt{x}}}}{\tc}
\end{eqnarray}
for $\tc$ a constant specifying the characteristic time-scale of the
process and the variation is taken only with respect to the driving
function $\mathsf{F}$.  In such a case the stationarity condition
reduces to (\ref{oc:scwsv22}) alone. Upon averaging the equation for
the value function (\ref{oc:value}) with respect to a probability
amplitude and using (\ref{oc:scwsv22}) we obtain the equation for the
auxiliary field $B$
\begin{eqnarray}
\label{}
\partial_{t}B\left(\mathtt{x},t\right)=\frac{2}{\tc}\,\sum_{\tilde{\mathtt{y}}\in\mathbb{S}}
\frac{\phi(\tilde{\mathtt{y}},t)}{\phi(\mathtt{x},t)}
\sinh\frac{\mathsf{F}(\tilde{\mathtt{y}},\mathtt{x},t)}{2}
\tanh\frac{\mathsf{F}(\tilde{\mathtt{y}},\mathtt{x},t)}{2}
\end{eqnarray}
Observing that
\begin{eqnarray}
\label{}
\partial_{t}[B\left(\mathtt{x},t\right)-B\left(\tilde{\mathtt{x}},t\right)]
=-\left[
2-\tanh^{2}\frac{\mathsf{F}(\mathtt{x},\tilde{\mathtt{x}},t)}{2}
\right]
\partial_{t}\mathsf{F}(\mathtt{x},\tilde{\mathtt{x}},t)
\end{eqnarray}
we get into the equations satisfied by the driving function
\begin{eqnarray}
\label{oc:psieq}
\lefteqn{
\partial_{t}\mathsf{F}(\mathtt{x},\tilde{\mathtt{x}},t)=
}
\nonumber\\&&
-\sum_{\mathtt{y}\in\mathbb{S}}
\frac{\frac{\phi(\mathtt{y},t)}{\phi\left(\mathtt{x},t\right)}\sinh\frac{\mathsf{F}(\mathtt{y},\mathtt{x},t)}{2}
\tanh\frac{\mathsf{F}(\mathtt{y},\mathtt{x},t)}{2}
-\frac{\phi(\mathtt{y},t)}{\phi\left(\tilde{\mathtt{x}},t\right)}
\sinh\frac{\mathsf{F}(\mathtt{y},\tilde{\mathtt{x}},t)}{2}
\tanh\frac{\mathsf{F}(\mathtt{y},\tilde{\mathtt{x}},t)}{2}}
{\tc\,\left[1-\frac{1}{2}\tanh^{2}\frac{\mathsf{F}(\mathtt{x},\tilde{\mathtt{x}},t)}{2}\right]
}
\end{eqnarray}
Degrees of freedom counting reveals, however, that not all transition
rates of the optimal process can be non-vanishing. Let us suppose
first $|\mathbb{S}|=N$ and then deduce the result for infinite lattice
from the limit $N\uparrow \infty$.  Because of probability
conservation (\ref{oc:pc}), the boundary conditions impose $2\,N-2$
independent conditions. By virtue of
$\mathsf{H}^{\dagger}=-\mathsf{H}$, the evolution of probability
amplitudes is probability preserving and brings forth $N-1$
independent equations.  We conclude that (\ref{oc:psieq}) can only
describe the dynamics of other $N-1$ degrees of freedom. We identify
these degrees of freedom by reasoning that $N-1$ independent
non-vanishing transition rates are exactly those needed to describe a
process jumping only to nearest neighbors states. If we posit that the
state space lattice spacing is $dx\,>\,0$, we achieve a convenient
parametrization of the nearest-neighbor dynamics if we define the
discrete \emph{current velocity}
\begin{eqnarray}
\label{}
V(\mathtt{x},t):=\mathsf{F}\left(\mathtt{x}+dx,\mathtt{x},t\right)
=-\mathsf{F}\left(\mathtt{x},\mathtt{x}+dx,t\right)
\end{eqnarray}
Gleaning the considerations expounded above we infer that the
transition between two assigned states releasing the minimal amount of
heat is specified by the following system of Hamilton-Jacobi-Bellman
\cite{FlemingSoner} and probability amplitude transport equations
\begin{subequations}
\label{oc:oceq}
\begin{eqnarray}
\label{oc:hjb}
\lefteqn{
\tc\,\partial_{t}V(\mathtt{x},t)=
}
\nonumber\\&&
\frac{2\,\sinh\frac{V(\mathtt{x},t)}{2}\tanh\frac{V(\mathtt{x},t)}{2}}{2-\tanh^{2}\frac{V(\mathtt{x},t)}{2}}
\left[\frac{\phi(\mathtt{x}+dx,t)}{\phi(\mathtt{x},t) }-
\frac{\phi(\mathtt{x},t)}{\phi(\mathtt{x}+dx,t)} \right]
\nonumber\\&&
+\frac{2\,\sinh\frac{V(\mathtt{x}-dx,t)}{2}\tanh\frac{V(\mathtt{x}-dx,t)}{2}}{2-\tanh^{2}\frac{V(\mathtt{x},t)}{2}}
\frac{\phi(\mathtt{x}-dx,t)}{\phi(\mathtt{x},t)}
\nonumber\\&&
-\frac{2\,\sinh\frac{V(\mathtt{x}+dx,t)}{2}\tanh\frac{V(\mathtt{x}+dx,t)}{2}}{2-\tanh^{2}\frac{V(\mathtt{x},t)}{2}}
\frac{\phi(\mathtt{x}+2\,dx,t)}{\phi(\mathtt{x}+dx,t)}
\end{eqnarray}
\begin{eqnarray}
\label{oc:pae}
\lefteqn{
\tc\,\partial_{t}\phi(\mathtt{x},t)=
}
\nonumber\\&&
-\left\{\sinh\frac{V(\mathtt{x},t)}{2}
\,\phi(\mathtt{x}+dx,t)-\sinh\frac{V(\mathtt{x}-dx,t)}{2}\,\phi(\mathtt{x}-dx,t)\right\}
\end{eqnarray}
\end{subequations}
with boundary conditions
\begin{eqnarray}
\label{oc:bc}
\phi\left(\mathtt{x},\ti\right)=\phi_{o}\left(\mathtt{x}\right)
\hspace{1.0cm}\&\hspace{1.0cm}
\phi\left(\mathtt{x},\tf\right)
=\phi_{f}\left(\mathtt{x}\right)
\end{eqnarray}
The value of the total entropy variation along the stationary control (\ref{oc:scwsv22})
\begin{eqnarray}
\label{oc:minev}
\mathcal{S}_{\star}=2\,\int_{t_{o}}^{\tf}\frac{dt}{\tc}\sum_{\mathtt{x}\in\mathbb{S}}
\,V(\mathtt{x},t)\,\sinh\frac{V(\mathtt{x},t)}{2}\,
\phi(\mathtt{x}+dx,t)\,\phi(\mathtt{x},t)
\end{eqnarray}
with the convention $\phi(\mathtt{y},t)=0$ if $\mathtt{y} \nin \mathbb{S}$ in case $|\mathbb{S}|\,<\,\infty$.

\subsection{Second variation in the presence of constrained reduced traffic}

We now show that the second variation of the total entropy with
respect to the driving function is positive definite around the
stationary point specified by (\ref{oc:scwsv22}) for fixed reduced
traffic. For any admissible control, we can write
\begin{eqnarray}
\label{oc:2v}
\lefteqn{
\mathcal{F}^{\prime\prime}_{\mathsf{F}}=
\left(\boldsymbol{\phi}\cdot\mathsf{J}^{\prime\prime}_{\mathsf{F}}\cdot\boldsymbol{\phi}\right)(\ti)
}
\nonumber\\&&
=\int_{\ti}^{\tf}dt\,\boldsymbol{\phi}\cdot\left(\mathsf{U}_{\mathsf{F}}^{\prime\prime}
-2\,[\mathsf{H}_{\mathsf{F}}^{\prime}\,,\mathsf{J}^{\prime}_{\mathsf{F}}]-
[\mathsf{H}_{\mathsf{F}}^{\prime\prime}\,,\mathsf{J}]\,\right)
\cdot\boldsymbol{\phi}
\end{eqnarray}
The first and third integrands in (\ref{oc:2v}) give positive definite
contributions to the second variation when evaluated at stationarity.
Namely
\begin{eqnarray}
\label{}
\lefteqn{
\frac{\partial^{2} \mathsf{U}(\mathtt{x},\tilde{\mathtt{x}})}
{\partial \mathsf{F}(\mathtt{x},\tilde{\mathtt{x}},t)
\partial \mathsf{F}(\mathtt{x},\tilde{\mathtt{x}},t)}=
}
\nonumber\\&&
\frac{1}{\tc}
\left[\cosh\frac{\mathsf{F}(\mathtt{x},\tilde{\mathtt{x}},t)}{2}
+\frac{\mathsf{F}(\mathtt{x},\tilde{\mathtt{x}},t)}{4}
\sinh\frac{\mathsf{F}(\mathtt{x},\tilde{\mathtt{x}},t)}{2}\right]\geq 0
\end{eqnarray}
and, since $\mathsf{H}_{\mathsf{F}}^{\prime\prime}$ is anti-symmetric, 
\begin{eqnarray}
\label{}
\lefteqn{
\boldsymbol{\phi}\cdot\left[\mathsf{H}^{\prime\prime}_{\mathsf{F}}\,,\mathsf{J}\right]\cdot\boldsymbol{\phi}=
\sum_{\mathtt{x},\tilde{\mathtt{x}}\in\mathbb{S}}\mathsf{H}^{\prime\prime}_{\mathsf{F}}(\mathtt{x},\tilde{\mathtt{x}},t)
\phi(\mathtt{x},t)\phi(\tilde{\mathtt{x}},t)\frac{B(\mathtt{x},t)-B(\tilde{\mathtt{x}},t)}{2}=
}
\nonumber\\&&
-\sum_{\mathtt{x},\mathtt{y}\in\mathbb{S}}\mathsf{H}^{\prime\prime}_{\mathsf{F}}(\mathtt{x},\tilde{\mathtt{x}},t)
\phi(\mathtt{x},t)\phi(\tilde{\mathtt{x}},t)\,\left\{
2\,\tanh\frac{\mathsf{F}(\mathtt{x},\tilde{\mathtt{x}},t)}{2}
+\mathsf{F}(\mathtt{x},\tilde{\mathtt{x}},t)
\right\}
\end{eqnarray}
whence it follows
\begin{eqnarray}
\label{}
\left.\boldsymbol{\phi}\cdot\left[\mathsf{H}^{\prime\prime}_{\mathsf{F}}\,,\mathsf{J}\right]\cdot\boldsymbol{\phi}
\right|_{\mathsf{X}^{\prime\prime}=0}\,\leq\, 0
\end{eqnarray}

Finally, we observe that 
\begin{eqnarray}
\label{oc:2vzero}
\int_{\ti}^{\tf}dt\,(\boldsymbol{\phi}\cdot\left[
\mathsf{H}_{\mathsf{F}}^{\prime}\,,\mathsf{J}_{\mathsf{F}}^{\prime}\right]\cdot\boldsymbol{\phi})(t)
=
0
\end{eqnarray}
since we can re-write it as the trace of the product of a symmetric and an anti-symmetric matrix. 
As a consequence we proved that
\begin{eqnarray}
\label{}
\mathcal{F}^{\prime\prime}_{\mathsf{F}_{\star}}=\int_{\ti}^{\tf}dt\,\boldsymbol{\phi}\cdot\left(\mathsf{U}_{\mathsf{F}}^{\prime\prime}
-[\mathsf{H}_{\mathsf{F}}^{\prime\prime}\,,\mathsf{J}]\,\right)
\cdot\boldsymbol{\phi}|_{\mathsf{F}_{\star}}\geq 0
\end{eqnarray}

\section{Continuum limit}
\label{sec:cl}

The limit of vanishing lattice spacing $dx$ for short distance
interactions is most straightforward to derive if we posit
\begin{eqnarray}
\label{cl:jump}
\mathsf{K}_{t}\left(x\pm dx|x\right)=\frac{1}{2\,\tc}
\left\{2\pm\,\beta\,b\left(x,t\right)\,dx\right\}
:=\frac{k_{\pm}(x,t)}{\tc}
\end{eqnarray}
We define the continuum limit of the Markov jump process with jump
rates (\ref{cl:jump}) by letting $dx$ tend to zero while fine-tuning
the characteristic time as
\begin{eqnarray}
\label{cl:diffusive}
\tc=\beta\,(dx)^{2}\,\tau
\end{eqnarray}
for some finite $\tau\,>\,0$ and measuring $\beta$ in units such that
$\beta dx^{2}$ is non-dimensional.  In such a case the limit
\begin{eqnarray}
\label{}
\lefteqn{
(\mathfrak{L}f)(x):=\lim_{dx\downarrow 0}(\mathsf{L}f)\left(x\right)=
}
\nonumber\\&&
\lim_{dx\downarrow 0}\frac{
[f\left(x+dx\right)-f\left(x\right)]
\,k_{+}\left(x,t\right)+[f\left(x-dx\right)-f\left(x\right)]
\,k_{-}\left(x,t\right)}{(dx)^{2}\,\tau}
\end{eqnarray} 
is finite and yields
\begin{eqnarray}
\label{}
\mathfrak{L}_{x}=\frac{b\left(x,t\right)}{\tau}\partial_{x}+\frac{1}{\beta\,\tau}\partial_{x}^{2}
\end{eqnarray}
which is the generator of the stochastic process described by the
stochastic differential equation
\begin{eqnarray}
\label{}
d\xi_{t}=b(\xi_{t},t)\,\frac{dt}{\tau}+\sqrt{\frac{2}{\beta\,\tau}}\,d\omega_{t}
\end{eqnarray}
driven by a Wiener process $\omega_{t}$.

\subsection{Heat density}

Under the hypotheses formulated above, the heat functional reduces to
\begin{eqnarray}
\label{}
\lefteqn{
\sum_{x,\tilde{x}\in\mathbb{S}}
\ln \frac{\mathsf{K}_{t}(x|\tilde{x})}{\mathsf{K}_{t}(\tilde{x}|x)}\,
\mathsf{K}_{t}(x|\tilde{x})=
}
\nonumber\\&&
 -\left[
\frac{\beta\,b(x,t)}{2}+\frac{\beta\,b(x+dx,t)}{2}
\right]\frac{2+b(x,t)\,dx}{2\,\beta\,(dx)^{2}\,\tau}
\nonumber\\&&
-\frac{1}{(dx)^{2}\,\tau}\left[-\frac{\beta\,b(x,t)}{2}-\frac{\beta\,b(x-dx,t)}{2}
\right]\frac{2-b(x,t)\,dx}{2\,\beta\,(dx)^{2}\,\tau}+O(dx)^{2}
\end{eqnarray}
whence finally we get into
\begin{eqnarray}
\label{}
\sum_{x,\tilde{x}\in\mathbb{S}}
\ln \frac{\mathsf{K}_{t}(x|\tilde{x})}{\mathsf{K}_{t}(\tilde{x}|x)}\,
\mathsf{K}_{t}(x|\tilde{x})=\frac{\beta\,b^{2}(x,t)-\partial_{x}b(x,t)}{\tau}
+O(dx)^{2}
\end{eqnarray}
We thus recover the expression of the heat density released by
Langevin dynamics see e.g. \cite{AuMeMG11}.  Thus, it is a-priori
justified to expect that (\ref{oc:oceq}) admits as continuum limit the
optimal transport equations found in \cite{AuMeMG11,AuMeMG12}.  In
order to derive explicitly this result, we define the continuous limit
current velocity $v$ by Taylor expanding its lattice counter-part in
powers of the mesh $dx$:
\begin{eqnarray}
\label{cl:cv}
\frac{V(x,t)}{\tau}\equiv\frac{\mathsf{F}(x+dx,x,t)}{\tau}=
v(x,t)dx+\frac{1}{2}\partial_{x}v(x,t)\,(dx)^{2}+O\left(dx\right)^{3}
\end{eqnarray}

\subsection{Probability density equation}

Upon inserting (\ref{cl:cv}) into (\ref{oc:pae}) and rescaling time
according to (\ref{cl:diffusive}), we obtain
\begin{eqnarray}
\label{}
\lefteqn{
\partial_{t}\phi(\mathtt{x},t)=
}
\nonumber\\&&
-\left\{\frac{dx\,v(\mathtt{x},t)+\frac{(dx)^{2}}{2}\partial_{x}v(x,t)}{2\,(dx)^{2}\,\tau}\right\}
\left\{\phi(x,t)+dx\,\partial_{x}\phi(x,t)\right\}
\nonumber\\&&
+\left\{\frac{dx\,v(\mathtt{x},t)-\frac{(dx)^{2}}{2}\partial_{x}v(x,t)}{2\,(dx)^{2}\,\tau}\right\}
\left\{\phi(x,t)-dx\,\partial_{x}\phi(x,t)\right\}+O(dx)
\end{eqnarray}
which yields 
\begin{eqnarray}
\label{cl:pae}
\partial_{t}\phi(x,t)=-\phi(x,t)\left\{v(x,t)\partial_{x}\phi(x,t)+\frac{1}{2}\partial_{x}v(x,t)\right\}
\end{eqnarray}
Multiplying (\ref{cl:pae}) by $2\,\phi$ recovers the probability
transport equation by the current velocity.

\subsection{Control equation}

Since
\begin{eqnarray}
\label{}
\partial_{t}\mathsf{F}(x+dx,x,t)=dx\,\partial_{t}v(x,t)+O(dx^{2})
\end{eqnarray}
the continuum limit of (\ref{oc:hjb}) calls for an expansion up to
third order in $dx$:
\begin{eqnarray}
\label{}
\lefteqn{
dx^{3}\,\tau\,(\partial_{t}v)\left(x,t\right)
=}
\nonumber\\&&
\hspace{0.33cm}\frac{1}{4}\,(dx)^{2}\,[v\left(x,t\right)
+\frac{dx}{2}\,(\partial_{x}v)\left(x,t\right)]^{2}\,2\,dx\,\partial_{x}\ln\phi(x,t)
\nonumber\\&&
+\frac{1}{4}\,(dx)^{2}\,[v\left(x,t\right)-\frac{dx}{2}
\,(\partial_{x}v)\left(x,t\right)]^{2}\left[1-dx\,\partial_{x}\ln\phi(x,t)\right]
\nonumber\\&&
-\frac{1}{4}\,(dx)^{2}\,[v\left(x,t\right)+3\frac{dx}{2}
\,(\partial_{x}v)\left(x,t\right)]^{2}\left[1+dx\,\partial_{x}\ln\phi(x,t)\right]+O(dx)^{4}
\end{eqnarray}
Unfolding the products we get into
\begin{eqnarray}
\label{}
\tau\,\partial_{t}v(x,t)+v(x,t)\partial_{x}v(x,t)=0
\end{eqnarray}
which is the equation for the current velocity found in
\cite{AuMeMG11,AuMeMG12}.

\section{Examples: two and three-state systems}
\label{sec:3s}

To simplify the notation it is expedient to measure time in units of the traffic rate, $\tc=1$,
and to consider a unit lattice spacing, $dx=1$.

\subsection{Two-state system: comparison with optimal control for Glauber transition rates }

In the case of a two-state jump process, probability conservation yields a closed single equation for 
the evolution of the probability amplitude of the first state
\begin{eqnarray}
\label{2s:pae}
\partial_{t}\phi(0,t)=-\sqrt{1-\phi(0,t)^{2}} \,
\sinh\frac{V\left(0,t\right)}{2}
\end{eqnarray}
We can, thus, proceed as in \cite{EsKaLiBr10} and write the entropy production (\ref{oc:minev}) 
as a functional of the occupation probability 
\begin{eqnarray}
\label{}
p(t)=\phi(0,t)^{2}
\end{eqnarray}
of the first state. We obtain
\begin{subequations}
\label{2s:epo}
\begin{eqnarray}
\label{}
\mathcal{S}=\int_{\ti}^{\tf}dt\,\sigma(p,\dot{p})
\end{eqnarray}
\begin{eqnarray}
\label{}
\sigma(x,y)=-y\,ln\frac{[\sqrt{4\,x\,(1-x)+y^{2}}-y]^{2}}{4\,x\,(1-x)}\geq 0
\end{eqnarray}
\end{subequations}
A straightforward and somewhat tedious calculation shows that the Euler-Lagrange equation 
specifying the trajectory of $p$ corresponding to an extremal point of (\ref{2s:epo}) coincides 
with the one obtained by differentiating (\ref{2s:pae}) and using
\begin{eqnarray}
\label{}
\partial_{t}V(0,t)
-\frac{2\,\sinh\frac{V(0,t)}{2}\tanh\frac{V(0,t)}{2}}{2-\tanh^{2}\frac{V(0,t)}{2}}
\left(\sqrt{ \frac{1-p}{p}}-\sqrt{ \frac{p}{1-p}} \right)
\end{eqnarray}
to write a closed expression for the evolution of the occupation probability of the first state
\begin{eqnarray}
\label{}
\ddot{p}=\frac{2\,(1-2\,p)\dot{p}^{2}}{8\,p\,(1-p)+\dot{p}^{2}}
\end{eqnarray}
This is a consequence of the well-known fact (see e.g. \cite{GuMo83}) that any system of Euler-Lagrange equations
is equivalent to a suitable deterministic control problem via Hamilton-Jacobi theory. In writing (\ref{2s:epo})
we fixed the reduced traffic to a constant, $p$-independent value as required by the optimal control equations
of section~\ref{sec:oc}. It is interesting to compare the entropy production in (\ref{2s:epo}) with the 
one corresponding to a Glauber jump process with transition rate
\begin{eqnarray}
\label{2s:Gr}
\mathsf{K}_{t}(\mathtt{0}|\mathtt{1})=\frac{1}{e^{h}+1}
\end{eqnarray}
since this is the jump process considered in \cite{EsKaLiBr10}. We obtain
\begin{subequations}
\label{2s:epogl}
\begin{eqnarray}
\label{}
\mathcal{S}_{Gl}=\int_{\ti}^{\tf}dt\,\sigma_{Gl}(p,\dot{p})
\end{eqnarray}
\begin{eqnarray}
\label{}
\sigma_{Gl}(x,y)=-y\,ln\frac{x\,(1-x-y)}{(1-x)\,(x+y)}\,\geq\,0
\end{eqnarray}
\end{subequations}
\begin{figure}
  \centering
\setlength{\unitlength}{0.1cm}
\begin{picture}{(0,0)}
\put(57,35){$\scriptstyle{x=0.05}$}
\put(60,27){$\scriptstyle{x=0.35}$}
\put(3,27){$\scriptstyle{x=0.65}$}
\put(22,22){$\scriptstyle{x=0.95}$}
\put(42,40){$\sigma_{Gl}-\sigma$}
\put(68,0){$y$}
\end{picture}
  \includegraphics[width=7.0cm]{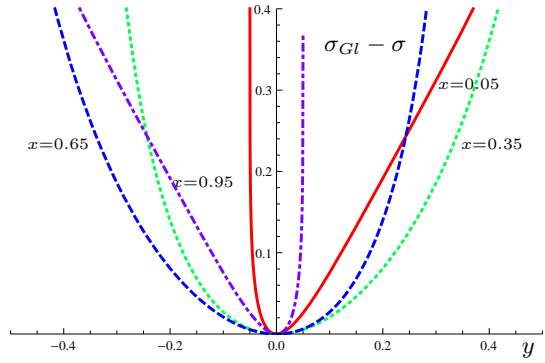}
  \caption{Numerical illustration of the inequality $(\sigma_{Gl}-\sigma)(x,y)\geq 0$. We plot the difference
   between the entropy production of the Glauber $\sigma_{Gl}$ and of the Hamilton-Jacobi-Bellman optimal $\sigma$ 
   jump processes for fixed value of the occupation probability $x\equiv p$ as a function of $y$. The continuous curve 
   corresponds to $x=0.05$, the dotted to $x=0.35$, the dashed to $x=0.65$ and the dash-dotted to $x=0.95$. 
   The inequality is proved analytically in the main text.
    }
  \label{fig:2senpro}
\end{figure}
A preliminary observation is that whilst (\ref{2s:epo}) is well-defined for any $0< x < 1$, (\ref{2s:epogl})
restricts admissible controls to those satisfying the additional condition $-x\,\leq \,y\leq 1-x$. 
Furthermore, the difference
\begin{eqnarray}
\label{2s:enprodiff}
(\sigma_{Gl}-\sigma)(x,y)=y\,\ln\frac{(x+y)\,[y^{2}+2\,x\,(1-x)-y\,\sqrt{ 4\,x\,(1-x)+y^{2}}]}{2\,x^{2}\,(1-x-y)}
\end{eqnarray}
is positive-definite for any  $0< x < 1$ and admissible value of $y$. To prove the claim it is sufficient show that
\begin{eqnarray}
\label{2s:ineq}
g(x,y):=\frac{(x+y)\,[y^{2}+2\,x\,(1-x)-y\,\sqrt{ 4\,x\,(1-x)+y^{2}}]}{2\,x^{2}\,(1-x-y)}\,\geq\,1
\end{eqnarray} 
if $y$ is positive while $0\leq g(x,y)< 1$ if $y$ is negative. Indeed (\ref{2s:ineq}) reduces to
\begin{eqnarray}
\label{}
y\left\{(y^{2}+x\,(2+y))-(x+y)\,\sqrt{4\,x\,(1-x)+y^{2}}\right\}\geq 0
\end{eqnarray}
the argument of the curly brackets being positive whenever
\begin{eqnarray}
\label{}
1-x-y+(x+y)^{2}\geq 0
\end{eqnarray}
The Glauber transition rate (\ref{2s:Gr}) corresponds to fixing the reduced traffic 
to 
\begin{eqnarray}
\label{}
\mathsf{G}(\mathtt{0},\mathtt{1})=\frac{1}{2\,\cosh\frac{h}{2}}
\end{eqnarray}
and the identification $\mathsf{A}(\mathtt{0},\mathtt{1})=-h$. These
modeling choices can be advocated with convincing physical
arguments. The existence of ``local'' optimal controls for special
choices of the reduced traffic suggests that the reduced traffic can
be consistently, and perhaps should be, more appropriately thought as
a configurational rather than a control parameter of a physical
system.

\subsection{Optimal control of the three-state system }

One of the simplest non-trivial application of the full-fledged
optimal transport equations (\ref{oc:oceq}) is to a system with three
states. The probability amplitude is governed by the equations
\begin{subequations}
\begin{eqnarray}
\label{}
\partial_{t}\phi(0,t)=-\phi(1,t) \,
\sinh\frac{V\left(0,t\right)}{2}
\end{eqnarray}
\begin{eqnarray}
\label{}
\partial_{t}\phi(2,t)=\phi(1,t) \,
\sinh\frac{V\left(1,t\right)}{2}
\end{eqnarray}
\end{subequations}
while the current velocity satisfies
\begin{subequations}
\label{3s:eqs}
\begin{eqnarray}
\label{}
\lefteqn{
\hspace{-1.0cm}
\partial_{t}V(0,t)
-\frac{2\,\sinh\frac{V(0,t)}{2}\tanh\frac{V(0,t)}{2}}{2-\tanh^{2}\frac{V(0,t)}{2}}
\left[\frac{\phi(1,t)}{\phi(0,t) }-
\frac{\phi(0,t)}{\phi(1,t)} \right]
}
\nonumber\\&&
+\frac{2\,\sinh\frac{V(1,t)}{2}\tanh\frac{V(1,t)}{2}}{2-\tanh^{2}\frac{V(0,t)}{2}}
\frac{\phi(2,t)}{\phi(1,t)} =0
\end{eqnarray}
\begin{eqnarray}
\label{}
\lefteqn{
\hspace{-1.0cm}
\partial_{t}V(1,t)
-\frac{2\,\sinh\frac{V(1,t)}{2}\tanh\frac{V(1,t)}{2}}{2-\tanh^{2}\frac{V(1,t)}{2}}
\left[\frac{\phi(2,t)}{\phi(1,t) }-
\frac{\phi(1,t)}{\phi(2,t)} \right]
}
\nonumber\\&&
-\frac{2\,\sinh\frac{V(0,t)}{2}\tanh\frac{V(0,t)}{2}}{2-\tanh^{2}\frac{V(1,t)}{2}}
\frac{\phi(0,t)}{\phi(1,t)}=0
\end{eqnarray}
\end{subequations}
with
\begin{eqnarray}
\label{}
\phi(1,t)=\sqrt{ 1-\phi^{2}(0,t)-\phi^{2}(2,t)}
\end{eqnarray}
The equations specify the minimal heat release protocol once we
specify the initial and final states through (\ref{oc:bc}).

\subsubsection{Numerical solution}

\begin{figure}
  \centering
\setlength{\unitlength}{0.1cm}
\begin{picture}{(0,0)}
\put(5,30){$\scriptstyle{m(0,t)}$}
\put(25,17){$\scriptstyle{m(1,t)}$}
\put(50,3){$\scriptstyle{0.05}$}
\put(45,0){$\scriptstyle{t}$}
\put(66,27){$\scriptstyle{v(0,t)}$}
\put(68,7){$\scriptstyle{v(1,t)}$}
\put(105,0){$\scriptstyle{t}$}
\end{picture}
  \subfloat[Probability]{\label{fig:prob}\includegraphics[width=5.0cm]{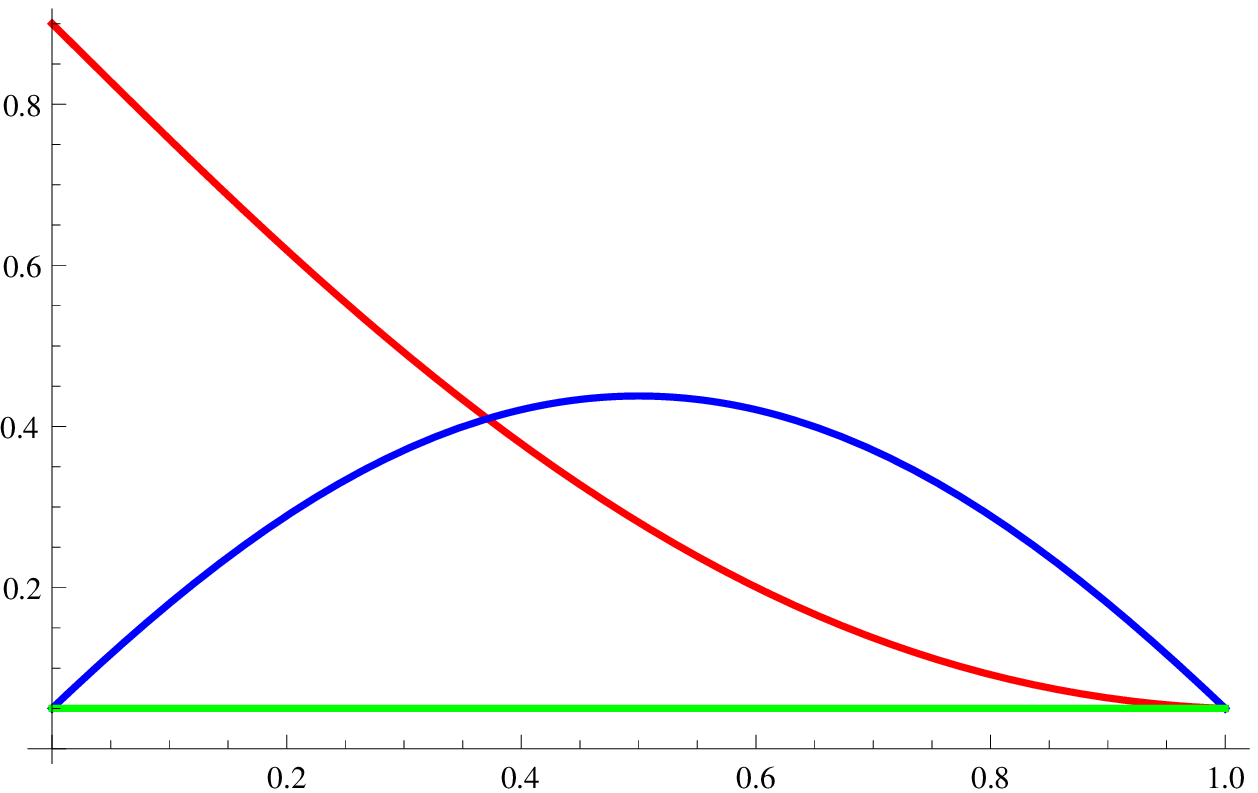}}
\hspace{1.0cm}
  \subfloat[Current velocity]{\label{fig:control}\includegraphics[width=5.0cm]{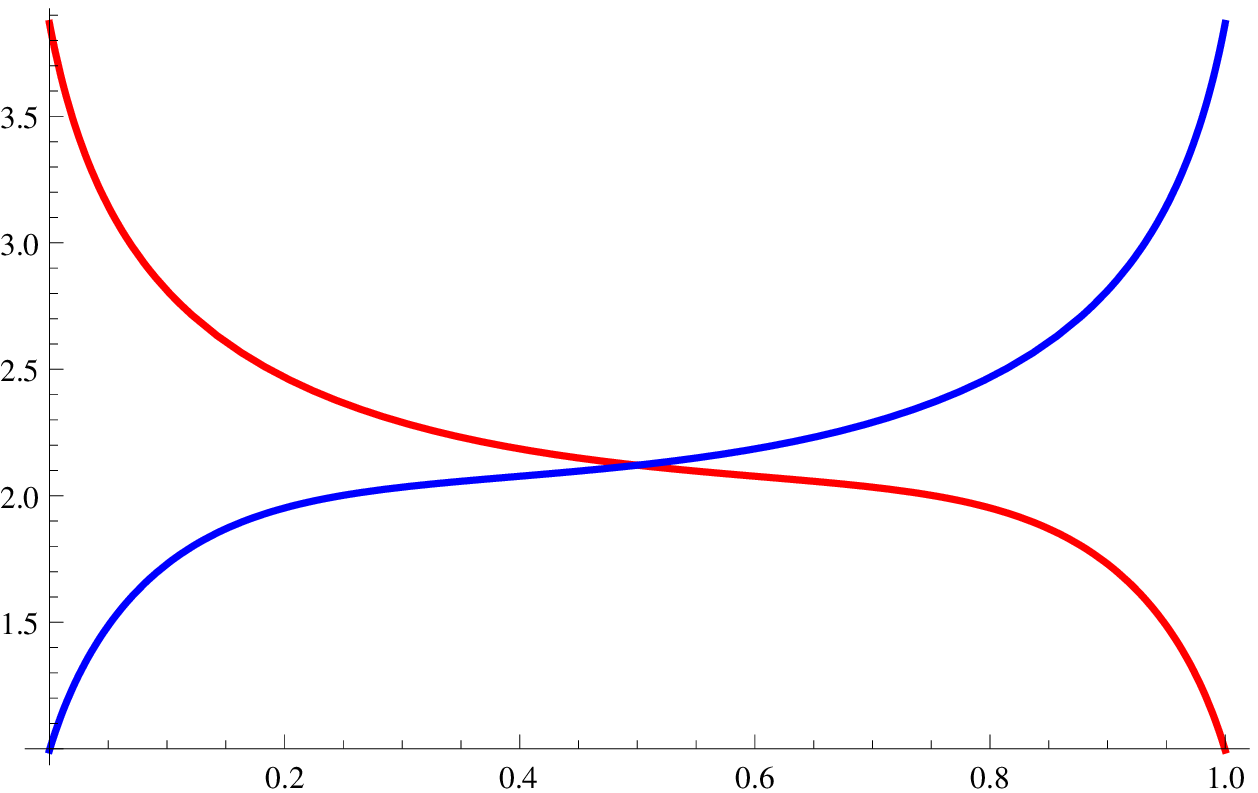}}
  \caption{Numerical solution of the three state process versus
    time. On the left (\ref{fig:prob}) the behavior of the occupation
    probability of the first $m(0,t)$ and the second $m(1,t)$
    state. On the right (\ref{fig:control}) the current velocity
    components $V(0,t)$, $V(1,t)$ driving the transition.}
  \label{fig:3snum}
\end{figure}

We considered the problem of minimizing the heat release during the 
transition between the states
\begin{eqnarray}
\label{}
\lefteqn{
\left[m(0,0),m(1,0),m(2,0)\right]=\left[0.9,0.05,0.05\right]\mapsto
}
\nonumber\\&&
\left[m(0,1),m(1,1),m(2,1)\right]=
\left[0.05,0.05,0.9\right]
\end{eqnarray}
We integrated numerically the system governing the evolution of the
occupation probabilities $[m(0,t),m(1,t)]$ and the discrete current
$[V(0,t),V(1,t)]$ using \emph{Wolfram's Mathematica (editions 7.0 and
  8.0.1)} \cite{Mathematica} default shooting method with at most
$4\times 10^{5}$ number of iterations. We used an initial Ansatz for
the pair $V(0,0),V(1,0)$ which we afterwards improved using the
candidate solutions produced by the shooting algorithm. The
improvement of the Ansatz shifted further in time the blow up of the
candidate solutions. After few manual iterations we obtained the pair
$[V_{\star}(0,0),V_{\star}(1,0)]=[3.150,1.156]$ which yields a smooth
solution over the full control horizon $[0,1]$ matching the boundary
conditions, see fig~\ref{fig:3snum}.  The entire procedure takes few
minutes using an old \emph{Pentium M} powered laptop.  We also checked
that the same results can be recovered using the probability amplitude
equations
but we noticed that this approach seems to increase the numerical stiffness of the problem. \\
The probability of the first state is always decreasing while the
probability of the intermediate state has a maximum at $t=0.5$ and
then symmetrically decays to match its boundary condition. The
constant line corresponds to $0.05$ and is plotted to emphasize the
convergence of the two occupation probabilities to the the assigned
boundary condition.  The optimal controls are fast varying near the
boundaries while varying more slowly in the bulk of the control
horizon. This is the characteristic behavior of optimal controls of
thermodynamics functionals described in
\cite{ScSe07,EsKaLiBr10,AuMeMG11,AuMeMG12,AuGaMeMoMG12}.  Since the Gibbs entropy
vanishes for this symmetric transition the heat release coincides with
the entropy production. The numerical evaluation of (\ref{oc:minev})
for the three state system then yields
\begin{eqnarray}
\label{}
\beta \mathcal{Q}_{\star}=\mathcal{S}_{\star}=4.096
\end{eqnarray}

\section{Conclusions}

We have shown that if the traffic is constrained to a constant value
the driving function governing the transition between two states with
minimal heat release in a finite time horizon obeys
Hamilton-Jacobi-Bellman type equations. The corresponding dynamics is
short-ranged as it permits only jumps to nearest-neighbor sites. Our
results are the counter-part for Markov jump processes of the entropy
bounds derived in \cite{AuGaMeMoMG12} for the Langevin dynamics. In
this sense, they can also be regarded as refinement of second law of
thermodynamics for Markov jump processes.  We have also shown that
entropy production is not convex in the traffic. As a consequence,
interpreting traffic as a control parameter results into
singular optimal control strategy. This means that any eventual heat
release occurs in a zero Lebesgue measure time interval. It is worth noticing
here that jump protocols solutions survive also in the Langevin continuum limit.
Namely, if we express as in \cite{AuMeMG12} the heat functional in terms of
the current velocity it is readily seen that the lower bound provided by 
the variation of the Gibbs-Shannon entropy becomes tight for a jump
protocol for which the current velocity is vanishing throughout the control 
horizon. This latter condition then just means that the drift must remain 
in equilibrium so that the resulting protocol is precisely the one found here
in the case of unconstrained traffic. In the Langevin case, this kind of
solutions for the heat functional can be ruled out a-priori by restricting
admissible control to \emph{smooth diffusions}. The current velocity representation also
evinces the subtle problems in which we incur when turning to 
thermodynamic work minimization. For the work it is not possible to find a minimizer
by restricting admissible protocols to smooth diffusions because of the 
non-coercive functional dependence on the control occasioned by the boundary cost term specified by
the internal energy variation.  Justifying in which sense variations for fixed terminal 
value of the internal energy may still define to an optimality criterion for 
a cost functional compatible with the interpretation of non-equilibrium thermodynamic work
calls then for a more refined analysis such as that carried out in \cite{AuMeMG12}.
An comprehensive discussion of these results can be found in \cite{PMG12}.

To summarize, we can say that the analysis of optimal control 
of Markov jump process helps to shed further light on the interpretation of the
result previously available for the Langevin dynamics. Restricting 
admissible controls to those determining the driving function 
provide then a simplified model for the continuous limit Langevin dynamics.
Such an approach is particularly beneficial
because, as exemplified by the three state jump process, numerical
solutions of the Hamilton-Jacobi-Bellman equations become extremely
computing time inexpensive in comparison to the
Monge-Amp\`ere-Kantorovich scheme solving the Langevin dynamics
\cite{AuGaMeMoMG12}. Still, the three state Markov jump process
captures relevant qualitative features of the Langevin dynamics. The
discrete dynamics is therefore very useful for capturing the
qualitative features of the optimal control of non-equilibrium
thermodynamics statistical indicators, such as higher moments of the
heat functional itself, the Hamilton-Jacobi-Bellman equations thereof
are not amenable, at least as far as it is currently known, to any
fast integration scheme in the Langevin limit.

\section{Acknowledgments}

It is a  pleasure to thank Erik Aurell  and Krzysztof Gaw\c{e}dzki for
many  enlightening   discussions  on  optimal   control  in  stochastic
thermodynamics.  P.M.-G.'s  work  was   supported  by  the  Center  of
Excellence  ``Analysis and  Dynamics''of the  Academy of  Finland. The
authors  gratefully  acknowledge  support  from the  European  Science
Foundation  and  hospitality  of  NORDITA  where this  work  has  been
initiated   during   their   stay   within  the   framework   of   the
”Non-equilibrium Statistical Mechanics” program. LP acknowledges the support 
of FARO and of PRIN 2009PYYZM5.

\appendix

\section*{Appendices}

\section{Mean backward derivative}
\label{ap:mbd}

Local detailed balance (\ref{jump:tr:ldb}) ensures that the
chain of equalities
\begin{eqnarray}
\label{}
\lefteqn{
\mathrm{E}_{\mathtt{x},t+dt}f(\xi_{t})=\sum_{\tilde{\mathtt{x}}\in\mathbb{S}}
f(\tilde{\mathtt{x}})\tilde{\mathsf{P}}_{t,t+dt}\left(\tilde{\mathtt{x}}\,|\mathtt{x}\right)}
\nonumber\\&&
=\sum_{\tilde{\mathtt{x}}\in\mathbb{S}}
\mathsf{P}_{t+dt,t}\left(\mathtt{x},t+dt\,|\,\tilde{\mathtt{x}},t\right)\tilde{\mathtt{x}}
\frac{m(\tilde{\mathtt{x}},t)}{m(\mathtt{x},t+dt)}
\end{eqnarray}
Expanding in Taylor series, neglecting terms $O(dt^{2})$ and
using the Master equations (\ref{jump:Master}), (\ref{jump:Master2})
yields the proof of (\ref{jump:tr:mbd}).

\section{Evaluation of the Kullback-Leibler divergence (\ref{jump:tr:KL})}
\label{ap:KL}

The first term on the right hand side of (\ref{jump:tr:KL}) 
is vanishing on average
\begin{eqnarray}
\label{}
\lefteqn{
\mathrm{E}^{(\xi)}\int_{\ti}^{\tf}dt\,
\sum_{\mathtt{x}\in\mathcal{S}}\left[\mathsf{K}_{t}(\mathtt{x}|\xi_{t})
-\bar{\mathsf{K}}_{t}(\mathtt{x}|\xi_{t})\right]
=
}
\nonumber\\&&
\int_{\ti}^{\tf}dt\,\sum_{\mathtt{x},\mathtt{y}\in\mathcal{S}}\left[\mathsf{K}_{t}(\mathtt{x}|\mathtt{y})-
\mathsf{K}_{t}(\mathtt{y}|\mathtt{x})\frac{\mathrm{m}(\mathtt{y},t)}{\mathrm{m}(\mathtt{x},t)}\right]
\mathrm{m}(\mathtt{x},t)=0
\end{eqnarray}
The second addend

\begin{eqnarray}
\label{}
\mathrm{E}^{(\xi)}\sum_{t\in\mathbb{J}(\xi)}\ln \frac{\mathsf{K}_{t}(\xi_{t}|\xi_{t_{-}})}
{\bar{\mathsf{K}}_{t}(\xi_{t}|\xi_{t_{-}})}\equiv
\int_{\ti}^{\tf}dt\sum_{\mathtt{x},\mathtt{y}\in\mathcal{S}}\ln \frac{\mathsf{K}_{t}(\mathtt{x}|\mathtt{y})
\,\mathrm{m}(\mathtt{y},t)}
{\mathsf{K}_{t}(\mathtt{y}|\mathtt{x})\,\mathrm{m}(\mathtt{x},t)}\mathsf{K}_{t}(\mathtt{x}|\mathtt{y})
\mathrm{m}(\mathtt{y},t)
\end{eqnarray}
carries two contributions. The first is the Shannon-Gibbs entropy of the
transformation
\begin{eqnarray}
\label{}
\lefteqn{
\int_{\ti}^{\tf}dt\sum_{\mathtt{x},\mathtt{y}\in\mathcal{S}}\ln \frac{\mathrm{m}(\mathtt{y},t)}
{\mathrm{m}(\mathtt{x},t)}\mathsf{K}_{t}(\mathtt{x}|\mathtt{y})
\mathrm{m}(\mathtt{y},t)
}
\nonumber\\&&
=\int_{\ti}^{\tf}dt\sum_{\mathtt{x},\mathtt{y}\in\mathcal{S}}\ln \mathrm{m}(\mathtt{y},t)\,[
\mathsf{K}_{t}(\mathtt{x}|\mathtt{y})\,\mathrm{m}(\mathtt{y},t)
-\mathsf{K}_{t}(\mathtt{y}|\mathtt{x})\,\mathrm{m}(\mathtt{x},t)]
\nonumber\\&&
=-\int_{\ti}^{\tf}dt\,\sum_{\mathtt{y}\in\mathcal{S}}\ln \mathrm{m}(\mathtt{y},t)\,\partial_{t}\mathrm{m}(\mathtt{y},t)=
-\mathrm{E}^{\xi}\ln\frac{\mathrm{m}_{f}(\xi_{\tf})}{\mathrm{m}_{o}(\xi_{\ti})}
\end{eqnarray}
by probability conservation eq. (\ref{statement:pc}).
The second specifies the thermodynamic heat and is given in the main text.

\section{Alternative treatment of the variational problem}
\label{ap:alter}

We may state the optimization problem directly for the heat functional
using as independent control the reduced traffic and the driving
functional $\mathsf{A}$:
\begin{eqnarray}
\label{}
\beta\,\mathcal{Q}=
\int_{\ti}^{\tf}dt^{\prime}\sum_{\mathtt{x},\tilde{\mathtt{x}}\in\mathcal{S}}
\mathsf{A}(\mathtt{x},\tilde{\mathtt{x}},t)
\,\mathsf{G}(\mathtt{x},\tilde{\mathtt{x}},t)\,e^{\frac{\mathsf{A}(\mathtt{x},\tilde{\mathtt{x}},t)}{2}}
\,m(\tilde{\mathtt{x}},t)
\end{eqnarray}
The corresponding value function in such a case is the solution of the
backward Kolmogorov equation
\begin{eqnarray}
\label{ap:value}
[(\partial_{t}+\mathsf{L})J](\tilde{\mathtt{x}},t)
+\sum_{\mathtt{x}\in\mathbb{S}}\mathsf{A}(\mathtt{x},\tilde{\mathtt{x}},t)
\,\mathsf{G}(\mathtt{x},\tilde{\mathtt{x}},t)\,e^{\frac{\mathsf{A}(\mathtt{x},\tilde{\mathtt{x}},t)}{2}}=0
\end{eqnarray}
where, as in the main text, we require the variation of $J$ to vanish
at $\tf$ as we regard $J(\cdot,\tf)$ as a functional of
$m(\cdot,\tf)$.  By Dynkin formula (see e.g. \cite{Klebaner}) we must
have
\begin{eqnarray}
\label{ap:Dynkin}
\beta\,\mathcal{Q}=\sum_{\tilde{\mathtt{x}}}\left\{m(\tilde{x},\ti)\,J(\tilde{\mathtt{x}},\ti)
-m(\tilde{x},\tf)\,J(\tilde{\mathtt{x}},\tf)\right\}
\end{eqnarray}
whence using the boundary conditions ti follows that
\begin{eqnarray}
\label{}
\beta\,\mathcal{Q}^{\prime}=\sum_{\tilde{\mathtt{x}}}m(\tilde{x},\ti)\,J^{\prime}(\tilde{\mathtt{x}},\ti)
\end{eqnarray}
The weak sense variation of (\ref{ap:value}) yields the conditions
\begin{subequations}
\begin{eqnarray}
\label{}
0=J(\mathtt{x},t)-J(\tilde{\mathtt{x}},t)+\mathsf{A}(\mathtt{x},\tilde{\mathtt{x}},t)
\end{eqnarray}
\begin{eqnarray}
\label{}
\lefteqn{
0=\left\{m(\mathtt{x},t)\,e^{\frac{\mathsf{A}(\mathtt{x},\tilde{\mathtt{x}},t)}{2}}\
+m(\tilde{\mathtt{x}},t)\,e^{-\frac{\mathsf{A}(\mathtt{x},\tilde{\mathtt{x}},t)}{2}}\right\}
\,\frac{J(\mathtt{x},t)-J(\tilde{\mathtt{x}},t)}{2}
}
\nonumber\\&&
+\left\{e^{\frac{\mathsf{A}(\mathtt{x},\tilde{\mathtt{x}},t)}{2}}\,m(\tilde{\mathtt{x}},t)
-\,e^{-\frac{\mathsf{A}(\mathtt{x},\tilde{\mathtt{x}},t)}{2}}\,m(\mathtt{x},t)\right\}
\nonumber\\&&
+\frac{\mathsf{A}(\mathtt{x},\tilde{\mathtt{x}},t)}{2}
\left\{e^{\frac{\mathsf{A}(\mathtt{x},\tilde{\mathtt{x}},t)}{2}}\,m(\tilde{\mathtt{x}},t)
+\,e^{-\frac{\mathsf{A}(\mathtt{x},\tilde{\mathtt{x}},t)}{2}}\,m(\mathtt{x},t)\right\}
\end{eqnarray}
\end{subequations}
Upon inserting the Ansatz
\begin{subequations}
\begin{eqnarray}
\label{}
\mathsf{A}(\mathtt{x},\tilde{\mathtt{x}},t)
=\mathsf{F}(\mathtt{x},\tilde{\mathtt{x}},t)+\ln m(\mathtt{x},t)-\ln m(\tilde{\mathtt{x}},t)
\end{eqnarray}
\begin{eqnarray}
\label{}
J(x,t)=B(x,t)-\ln m(\mathtt{x},t)
\end{eqnarray}
\end{subequations}
we recover the stationary point equations (\ref{oc:scwsv2}).
 
\addcontentsline{toc}{section}{Bibliography}

\bibliography{/home/paolo/RESEARCH/BIBTEX/jabref}{} 
\bibliographystyle{myhabbrv}

\end{document}